\documentclass[slac_one]{revtex4}
\usepackage{graphicx}
\usepackage{fancyhdr}
\newcommand{\lsim}
{\mathrel{\raisebox{-.3em}{$\stackrel{\displaystyle <}{\sim}$}}}
\newcommand{\gsim}
{\mathrel{\raisebox{-.3em}{$\stackrel{\displaystyle >}{\sim}$}}}
\def\asymp#1%
{\mathrel{\raisebox{-.4em}{$\widetilde{\scriptstyle #1}$}}}

\def\Nequal#1%
{\mathrel{\raisebox{-.5em}{$\stackrel{=}{\scriptstyle\rm#1}$}}}
\newcommand{\dsl}[1]{\not \hspace{-0.7mm}#1}
\def\dsl{\mathpalette\make@slash}
\def\make@slash#1#2{\setbox\z@\hbox{$#1#2$}%
  \hbox to 0pt{\hss$#1/$\hss\kern-\wd0}\box0}

\def\beq{\begin{equation}}
\def\eeq{\end{equation}}
\def\beqar{\begin{eqnarray}}
\def\eeqar{\end{eqnarray}}
\def\barr#1{\begin{array}{#1}}
\def\earr{\end{array}}
\def\bfi{\begin{figure}}
\def\efi{\end{figure}}
\def\btab{\begin{table}}
\def\etab{\end{table}}
\def\bce{\begin{center}}
\def\ece{\end{center}}

\def\text{\textstyle}

\def\al{\alpha}

\def\ga{\gamma}
\def\de{\delta}

\def\reffi#1{\mbox{Figure~\ref{#1}}}

\def\citere#1{\mbox{Ref.~\cite{#1}}}
\def\citeres#1{\mbox{Refs.~\cite{#1}}}

\newcommand{\GeV}{\unskip\,\mathrm{GeV}}

\newcommand{\pba}{\unskip\,\mathrm{pb}}

\newcommand{\M}{{\cal{M}}}

\def\mathswitchr#1{\relax\ifmmode{\mathrm{#1}}\else$\mathrm{#1}$\fi}

\newcommand{\PW}{\mathswitchr W}

\newcommand{\PZ}{\mathswitchr Z}

\newcommand{\PH}{\mathswitchr H}
\newcommand{\Pe}{\mathswitchr e}

\newcommand{\Pne}{\mathswitch \nu_{\mathrm{e}}}

\newcommand{\Pd}{\mathswitchr d}

\newcommand{\Pubar}{\bar{\mathswitchr u}}

\newcommand{\Pep}{\mathswitchr {e^+}}
\newcommand{\Pem}{\mathswitchr {e^-}}

\def\mathswitch#1{\relax\ifmmode#1\else$#1$\fi}

\newcommand{\MW}{\mathswitch {M_\PW}}

\newcommand{\MH}{\mathswitch {M_\PH}}

\newcommand{\GW}{\Gamma_{\PW}}

\newcommand{\Whizard}{{\sc Whizard}}
\newcommand{\Madgraph}{{\sc Madgraph}}

\newcommand{\CompAZ}{{\sc CompAZ}}

\newcommand{\ggffff}{\gamma\gamma\to 4f}

\newcommand{\eewwffff}{\Pep\Pem\to\PW\PW\to 4f}

\newcommand{\ggffffg}{\ggffff\ga}

\newcommand{\ggtoww}{\gamma\gamma\to \PW\PW}

\newcommand{\ggwwffff}{\gamma\gamma\to \PW\PW\to 4f}

\def\solid{\raise.9mm\hbox{\protect\rule{1.1cm}{.2mm}}}
\def\dash{\raise.9mm\hbox{\protect\rule{2mm}{.2mm}}\hspace*{1mm}}

\begin{document}

\title{{\small{2005 International Linear Collider Workshop - Stanford,
U.S.A.}}\\ 
\vspace{12pt}
\boldmath{
Precision calculations for
$\gamma\gamma\to\PW\PW\to4\,\mathrm{fermions}\,(+\gamma)$
}} 

%

\author{A.\ Bredenstein, S.\ Dittmaier, M.\ Roth}
\affiliation{Max-Planck-Institut f\"ur Physik,
D-80805 M\"unchen, Germany}

\begin{abstract}
The ${\cal O}(\al)$ electroweak radiative corrections to $\ggwwffff$ within
the electroweak Standard Model are calculated
in double-pole approximation (DPA).
Virtual corrections are treated in DPA, 
and real-photonic corrections are based on complete lowest-order
matrix elements for $\ggffff{+}\ga$. 
The radiative corrections are implemented in a Monte Carlo generator
called {\sc Coffer$\gamma\gamma$}%
\footnote{The computer code can be obtained from the authors upon request.}%
, which optionally includes anomalous
triple and quartic gauge-boson couplings in addition and performs a
convolution over realistic spectra of the photon beams.
A brief survey of numerical results comprises
${\cal O}(\al)$ corrections to
integrated cross sections as well as to
angular and invariant-mass distributions.
\end{abstract}

\maketitle

\thispagestyle{fancy}
\addtocounter{footnote}{1}


\section{INTRODUCTION}

As an option at a future $\Pep\Pem$ linear collider, 
a photon (or $\ga\ga$) collider \cite{Ginzburg:1981vm}
found considerable interest in recent years.
It could provide us with information about new physics phenomena,
such as properties of Higgs bosons or of new particles,
which is in many respects complementary in the
$\Pep\Pem$ and $\ga\ga$ modes
(see, e.g., \citeres{Ginzburg:1981vm,DeRoeck:2003gv} and references therein).
Moreover, a $\ga\ga$ collider is a true W-boson-pair factory,
owing to the extremely high W-pair cross section, which tends to a
constant of about $80\pba$ in the high-energy limit (in the absence
of phase-space cuts), opening the possibility of precision studies in the
sector of electroweak gauge bosons. 
For instance, an analysis of anomalous gauge-boson couplings
in $\ggwwffff$ provides direct information on the $\gamma\PW\PW$ and
$\gamma\gamma\PW\PW$ interactions without interference from the
$\PZ$-boson sector. 
Either way, whether one is
interested in W-boson precision physics or in the search for new phenomena,
precise predictions for W-pair production are indispensable for signal
and background studies.
\looseness -1

As described in \citeres{Bredenstein:2004ef,Bredenstein:2005zk} in detail,
we have constructed a Monte Carlo generator
called {\sc Coffer$\gamma\gamma$} for $\ggffff$ and $\ggffffg$, which
particularly focusses on precise predictions for W-pair-mediated final 
states. Anomalous 
$\gamma\PW\PW$, $\gamma\gamma\PW\PW$, and $\gamma\gamma\PZ\PZ$
gauge-boson couplings as well as a loop-induced $\gamma\gamma\PH$ coupling
are optionally included in lowest-order predictions for $\ggffff$.
Moreover, electroweak radiative corrections of ${\cal O}(\alpha)$
are applied to processes $\ggwwffff$ in the so-called 
``double-pole approximation'' (DPA).
In the following we briefly describe the salient features of 
{\sc Coffer$\gamma\gamma$} and show some sample results.
More details and results (also for effects of anomalous couplings)
can be found in \citeres{Bredenstein:2004ef,Bredenstein:2005zk}.
\looseness -1

\section{LOWEST-ORDER PREDICTIONS}

In \citere{Bredenstein:2004ef} we have 
constructed a Monte Carlo event generator for lowest-order predictions
based on complete matrix elements for $\ggffff$~%
\footnote{Results for cross sections based on the full set of
$\ggffff$ diagrams were also presented in \citere{Moretti:1996nv}. 
However, no convolution over a realistic photon beam spectrum was
performed there.}
and $\ggffffg$. 
The final-state fermions are treated as massless, but
all $4f$ final states are supported.
For $4f$ production all helicity amplitudes (with and without
anomalous couplings) are explicitly given
as compact expressions in terms of spinor products.
Moreover, the introduction of finite gauge-boson decay widths and the
issue of gauge invariance are discussed carefully.
The possibility to convolute the cross sections with
realistic photon beam spectra is offered upon using the
parametrization of {\CompAZ} \cite{Zarnecki:2002qr}.
The Standard Model predictions were successfully compared
to results obtained with the multi-purpose packages
{\Whizard} \cite{Kilian:2001qz} and {\Madgraph} \cite{Stelzer:1994ta}.
\looseness -1

\begin{figure}
\setlength{\unitlength}{1cm}
\centerline{
\begin{picture}(7,6.9)
\put(-1.2,-13.1){\includegraphics{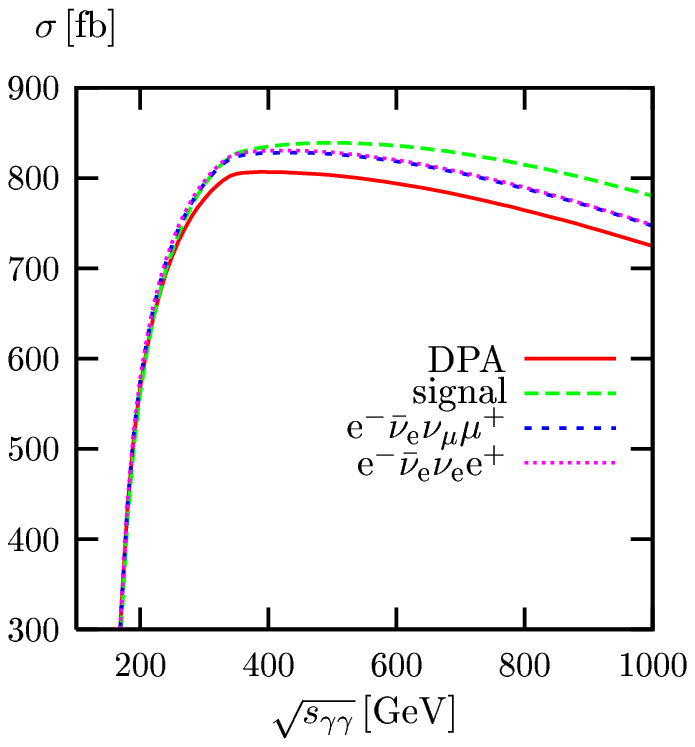}}
\end{picture}\hspace*{1em}
\begin{picture}(7,6.9)
\put(-1.2,-13.1){\includegraphics{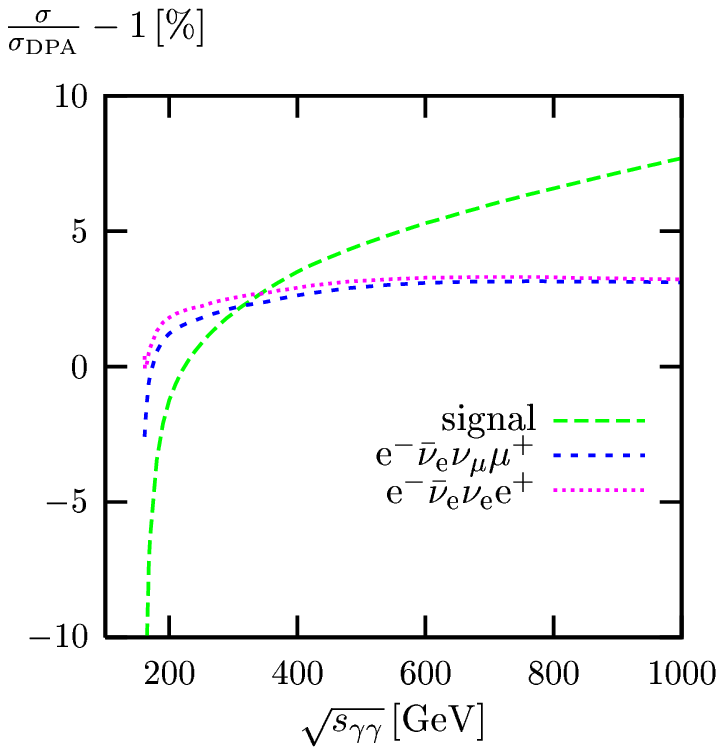}}
\end{picture}}
\vspace*{-.5em}
\caption{Lowest-order cross sections for
$\ga\ga\to\Pem\bar\nu_\Pe\nu_\mu\mu^+/\Pem\bar\nu_\Pe\nu_\Pe\Pep$
including all diagrams, only W-pair
signal diagrams, and in DPA as a function of the centre-of-mass (CM) energy 
$\sqrt{s_{\ga\ga}}$ of the monochromatic photon beams (l.h.s.),
and the corresponding relative deviations from the DPA (r.h.s.).
(Taken from \citere{Bredenstein:2004ef}.)}
\label{fig:locs}
\end{figure}

Among the various numerical results shown in \citere{Bredenstein:2004ef},
we only highlight the comparison between results that are based on the full
set of tree diagrams on the one hand and on the so-called ``signal'' diagrams
that involve two resonant W~bosons in $\ggwwffff$ on the other hand. 
Of course, the amplitude
for this ``naive W-pair signal'' is not a gauge-invariant quantity.
Nevertheless, its investigation is interesting, 
because such an amplitude is much simpler than the full
amplitudes for $4f$ production and is universal (up to colour factors)
for all relevant final states.
The DPA for the Born amplitude
is obtained from the naive W-pair signal upon deforming the
momenta of the four outgoing fermions in such a way that the
intermediate W-boson states become on shell, while keeping the W~propagators 
off shell.
Since the residues of the W~resonances are gauge independent,
the Born amplitude in DPA is a gauge-invariant quantity.
In \reffi{fig:locs}
the cross sections of the W-pair signal diagrams and the
DPA for $\ga\ga\to\PW\PW\to 4\,$leptons
are compared with the complete lowest-order cross sections.
The plot on the l.h.s.\ shows the cross sections, 
the plot on the
r.h.s.\ the relative deviation from the corresponding DPA.
We do not include the convolution over the
photon spectrum in this analysis
so that effects of the approximations are clearly visible.
For energies not too close to the W-pair threshold,
the DPA agrees with the full lowest-order cross section within
1--3\%, which is of the expected order of $\GW/\MW$.
Near threshold, i.e.\ for $\sqrt{s_{\ga\ga}}-2\MW={\cal O}(\GW)$,
the reliability of the DPA breaks down, since background
diagrams become more and more important and small scales $\ga$,
such as $\sqrt{s_{\ga\ga}-4\MW^2}$, can increase the naive error
estimate from $\GW/\MW$ to $\GW/\ga$.
The cross section of the W-pair signal diagrams, however, shows large
deviations from the full $\ggffff$ cross sections for the whole
energy range, in particular, at high energies. 
The results of \reffi{fig:locs} show that a naive
signal definition is a bad concept for $\ggwwffff$, since deviations
from the full process $\ggffff$
even reach 5--10\% in the TeV range. This is in contrast to the
situation at $\Pep\Pem$ colliders where the naive W-pair signal
(defined in `t~Hooft--Feynman gauge) was a
reasonable approximation
(see, e.g., \citere{Grunewald:2000ju}).

\section{RADIATIVE CORRECTIONS}

\subsection{Strategy of the calculation}

In \citere{Bredenstein:2005zk} we have extended 
our lowest-order calculation \cite{Bredenstein:2004ef}
for $\ggffff$ by including the electroweak radiative corrections of
${\cal O}(\alpha)$ to the W-pair channels $\ggwwffff$ in DPA.
The DPA extracts those contributions of the ${\cal O}(\alpha)$
corrections that are enhanced by two resonant W-boson
propagators, i.e.\ it represents the leading term in an expansion of
the cross section about the two W-propagator poles.
Note that tree-level diagrams for $\ggffff$ with at most one
resonant W~boson are suppressed w.r.t.\ the doubly-resonant
$\ggtoww$ signal by a factor of ${\cal O}(\GW/\MW)\sim {\cal O}(\alpha)$.
Consequently, predictions based on full lowest-order matrix elements for
$\ggffff$ and ${\cal O}(\alpha)$ corrections for $\ggwwffff$ in DPA
should be precise up to terms of ${\cal O}(\alpha/\pi\times \GW/\MW)$,
since corrections typically involve the factor $\alpha/\pi$.
Including a quite conservative numerical safety factor, the relative
uncertainty should thus be $\lsim0.5\%$ for such predictions,
as long as neglected effects are not additionally enhanced.
The naive error estimate can, in particular, be spoiled by the occurrence of
large scale ratios, which exist, e.g., near production thresholds or
at very high energies. The estimate has recently been confirmed
for $\eewwffff$ with CM energies
$170\GeV\lsim\sqrt{s}\lsim300\GeV$ by comparing a full
${\cal O}(\alpha)$ calculation \cite{Denner:2005es}
with the corresponding DPA predictions provided by {\sc RacoonWW}
\cite{Denner:1999gp}.

For energies in the W-pair threshold region ($\sqrt{s_{\ga\ga}}<170\GeV$)
and below, the DPA is unreliable, because diagrams with at most one
resonant W~boson become equally important. Therefore, in this region
we employ an ``improved Born approximation'' (IBA) which is based on
leading universal corrections but does not involve any resonance expansion.
As also confirmed by the full ${\cal O}(\alpha)$ calculation 
\cite{Denner:2005es} for the related process $\eewwffff$, such an IBA
reproduces the ${\cal O}(\alpha)$-corrected cross section typically
within $\sim\pm 2\%$. 
The convolution of the hard $\ga\ga$ cross section
involves both the IBA (in the low-energy tail) and the DPA
(for $\sqrt{s_{\ga\ga}}>170\GeV$).
As shown in \citere{Bredenstein:2005zk},
for CM energies (of the electrons before the Compton backscattering)
$\sqrt{s_{\Pe\Pe}}\lsim 230\GeV$ our prediction possesses an uncertainty
of $\sim 2\%$, because it is mainly based on the IBA, but already
for $\sqrt{s_{\Pe\Pe}}\gsim 300\GeV$ ($500\GeV$)
the IBA contribution is widely suppressed
so that the DPA uncertainty 
sets the precision of $\lsim 0.7\%$ ($0.5\%$) in our calculation.

In detail, we apply the DPA only to the virtual corrections to
$\ggwwffff$, while we base the real-photonic corrections on complete
lowest-order matrix elements for $\ggffffg$. Apart from the treatment
of IR (soft and collinear) singularities, we can use
the calculation of the bremsstrahlung
processes $\ggffffg$ for massless fermions
described in \citere{Bredenstein:2004ef}.
The concept of the DPA was already described in \citere{Aeppli:1993rs}
for the corrections to $\eewwffff$ and later successfully
applied to these processes in different versions
\cite{Denner:1999gp,Jadach:1998tz,Beenakker:1998gr,%
Kurihara:2001um}.
We follow the strategy of {\sc RacoonWW} \cite{Denner:1999gp}
and adapt it to $\gamma\gamma$ collisions where necessary.
The virtual corrections in DPA can be naturally split into
factorizable and non-factorizable contributions.
The former comprise the corrections to on-shell W-pair
production \cite{Denner:1994ca,Jikia:1996uu}
and the decay \cite{Bardin:1986fi} of on-shell W~bosons.
The latter account for soft-photon exchange between the production
and decay subprocesses; the known results for the non-factorizable
corrections \cite{Melnikov:1995fx}
for $\eewwffff$ can be taken over to $\gamma\gamma$ collisions with
minor modifications.

The combination of virtual and real-photonic corrections is non-trivial
for two reasons. First, the finite-fermion-mass effects have to be
restored in the phase-space regions of collinear photon radiation off
charged fermions, and the IR regularization for soft-photon emission
has to be implemented. To this end, we employ the dipole subtraction
formalism for photon radiation \cite{Dittmaier:1999mb}
as well as the
more conventional phase-space slicing approach.
The second subtlety concerns the fact that we apply the DPA only to
the virtual corrections, but not to the real-photonic parts.
Therefore, the cancellation of soft and collinear singularities
has to be done carefully, in order to avoid mismatch.

Finally, the Higgs-boson resonance in the $s$-channel
has to be treated with care. Firstly, the Higgs decay width has to
be introduced in the amplitude without violating gauge invariance.
We separate the gauge-invariant resonance pole and
include the width only in the resonant part which is proportional
to its gauge-invariant residue. Secondly, although the Higgs
resonance is loop-induced, it is not sufficient to include the
interference between its contribution $\M_{\mathrm{Higgs}}$ to the 
amplitude with the Born amplitude, but it is necessary to include the 
square $|\M_{\mathrm{Higgs}}|^2$ in the squared amplitude, in order
to get a proper description of the resonance.

\subsection{Numerical results}

\begin{figure}
\setlength{\unitlength}{1cm}
\centerline{
\begin{picture}(7.7,8)
\put(-1.7,-14.5){\includegraphics{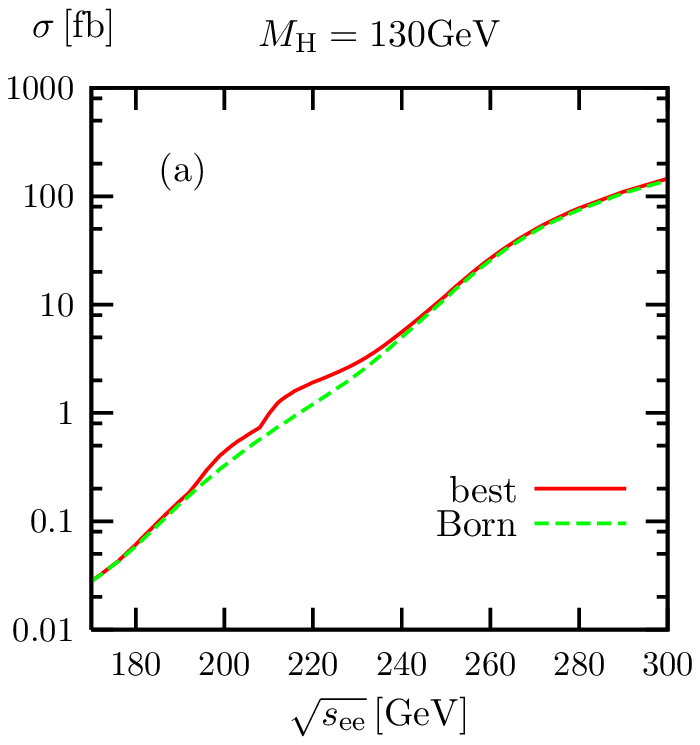}}
\end{picture}
\begin{picture}(7.5,8)
\put(-1.7,-14.5){\includegraphics{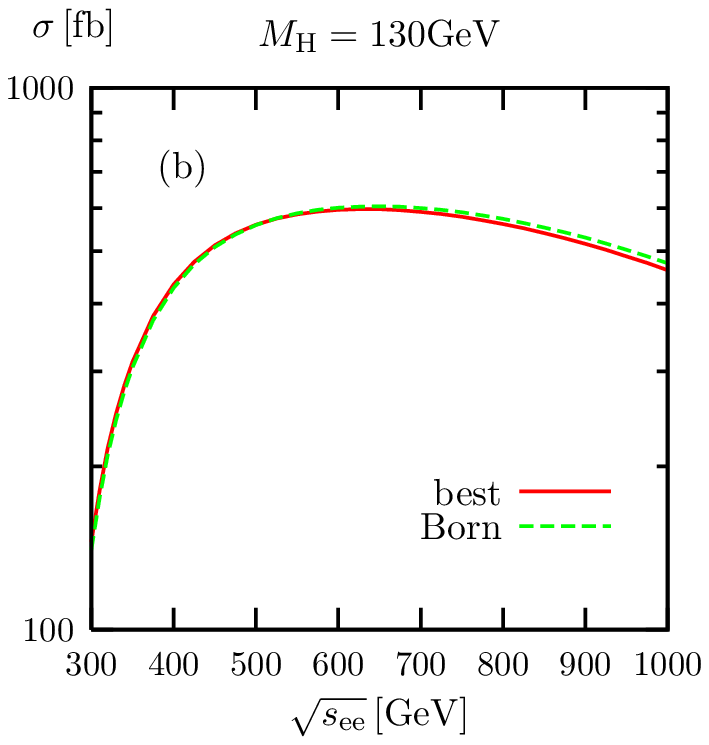}}
\end{picture} }
\vspace*{-.5em}
\centerline{
\begin{picture}(7.7,8)
\put(-1.7,-14.5){\includegraphics{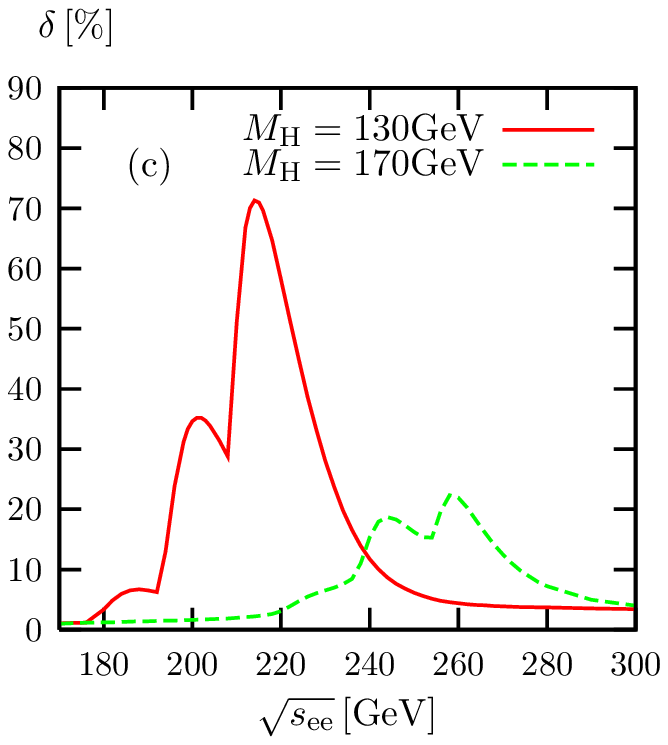}}
\end{picture}
\begin{picture}(7.5,8)
\put(-1.7,-14.5){\includegraphics{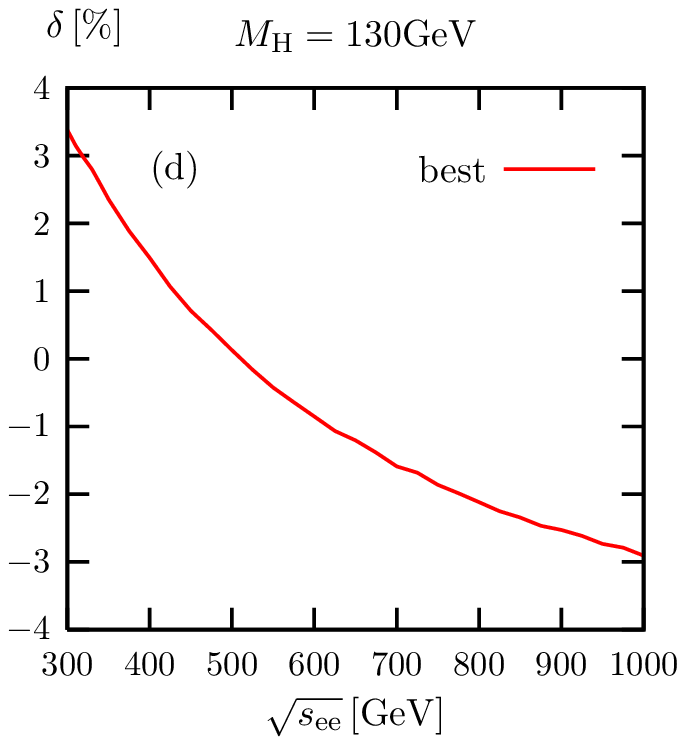}}
\end{picture} }
\vspace*{-1em}
\caption{Integrated cross section for $\ga\ga\to\Pne\Pep\Pd\Pubar$
(upper plots) and relative radiative
corrections (lower plots) including the convolution over the photon spectrum
for $\MH=130\GeV$ and $170\GeV$ (l.h.s.).
For $\sqrt{s_{\Pe\Pe}}>300\GeV$ (r.h.s.) the ``best'' curve
for $\MH=170\GeV$ practically coincides with the shown curve for
$\MH=130\GeV$. (Taken from \citere{Bredenstein:2005zk}.)}
\label{fig:sqrts.spec}
\end{figure}

The cross section for $\ga\ga\to\Pne\Pep\Pd\Pubar$,
including the convolution over the photon spectrum,
is shown in \reffi{fig:sqrts.spec}
as a function of CM energy $\sqrt{s_{\Pe\Pe}}$
for a Higgs mass of $\MH=130\GeV$ and in the lower left plot also
for $\MH=170\GeV$.
The remaining input parameters as well as the precise setup of the calculation are given in \citere{Bredenstein:2005zk}.
In the upper plots the absolute
prediction is shown and in the lower plots the corrections relative
to the Born cross section. 
The interesting structure in the lower left plot
reflects the shape of the photon spectrum convoluted with the Higgs resonance.
Since the Higgs resonance
is very narrow, a sizable contribution is only possible if
$x_1x_2s_{\Pe\Pe}\approx\MH^2$ where $x_1$ and $x_2$ are the energy fractions
carried by the photons. The correction is
very small at low $\sqrt{s_{\Pe\Pe}}$ where $x_1$ and $x_2$ both
have to be so large in order to match this condition
that the corresponding spectrum is extremely small. Increasing
$\sqrt{s_{\Pe\Pe}}$ allows for lower values of $x_1$ and $x_2$.
For instance, for $\MH=130\GeV$,
the rise at $\sqrt{s_{\Pe\Pe}}\sim 180\GeV$ results from a region
where both $x_1$ and $x_2$ are in the high-energy tail
of the spectrum which is produced by multiple photon scattering.
The peak at  $\sqrt{s_{\Pe\Pe}}\sim 200\GeV$ is caused by events where
one photon comes from the high-energy tail and one from the dominant
peak in the photon spectrum. Finally, at $\sqrt{s_{\Pe\Pe}}\gsim 210\GeV$
both $x_1$ and $x_2$ originate from
the dominant photon-spectrum peak which causes the steep rise
until $\sqrt{s_{\Pe\Pe}}\sim 220\GeV$.
For energies above the Higgs resonance [see \reffi{fig:sqrts.spec}(d)]
the corrections decrease and
stay in the range of a few per cent up to TeV energies.

\begin{figure}
\setlength{\unitlength}{1cm}
\centerline{
\begin{picture}(7.7,8)
\put(-1.7,-14.5){\includegraphics{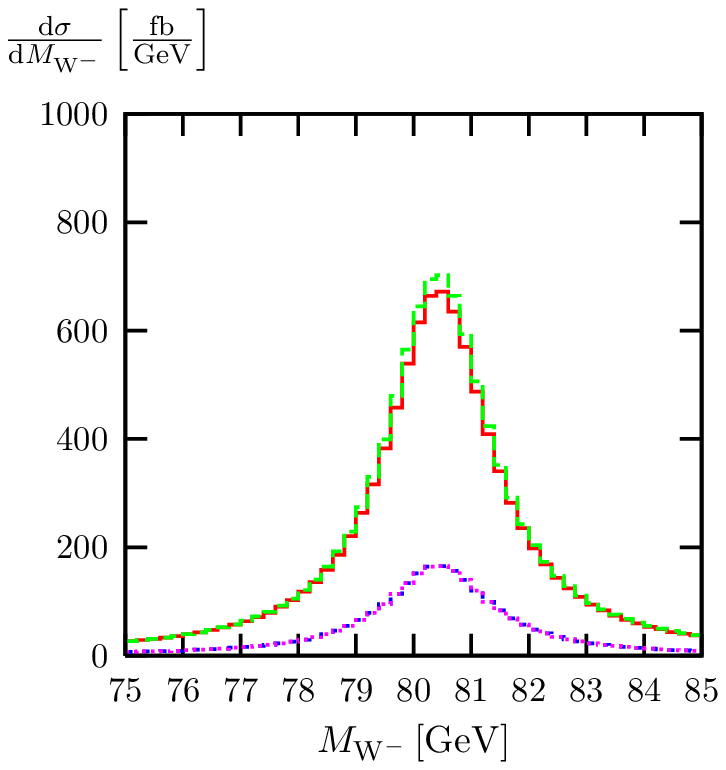}}
\end{picture}
\begin{picture}(7.5,8)
\put(-1.7,-14.5){\includegraphics{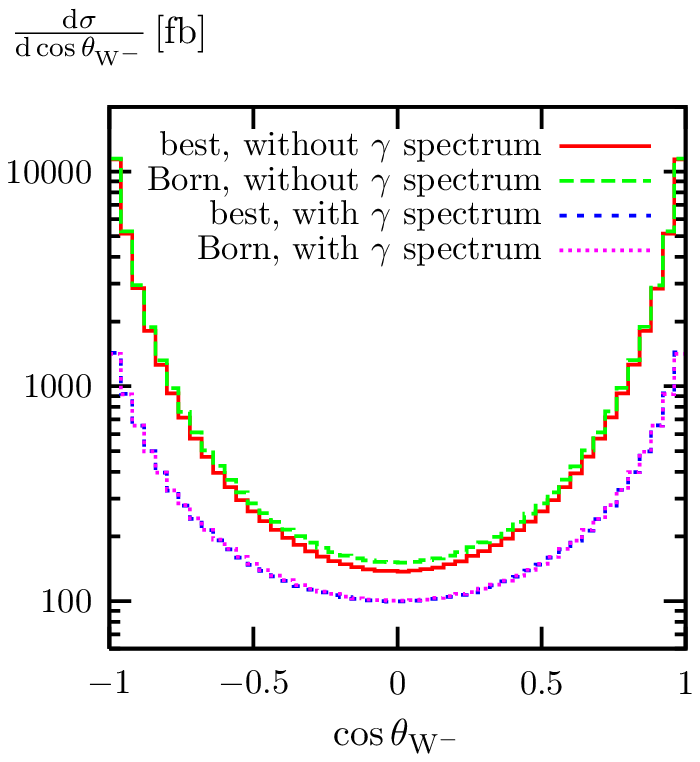}}
\end{picture}}
\vspace*{-1.2em}
\centerline{
\begin{picture}(7.7,8)
\put(-1.7,-14.5){\includegraphics{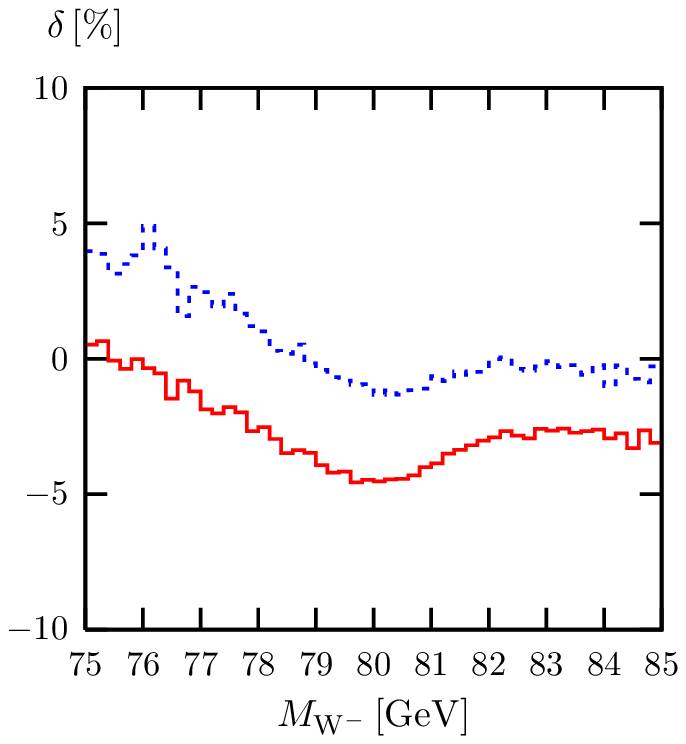}}
\end{picture}
\begin{picture}(7.5,8)
\put(-1.7,-14.5){\includegraphics{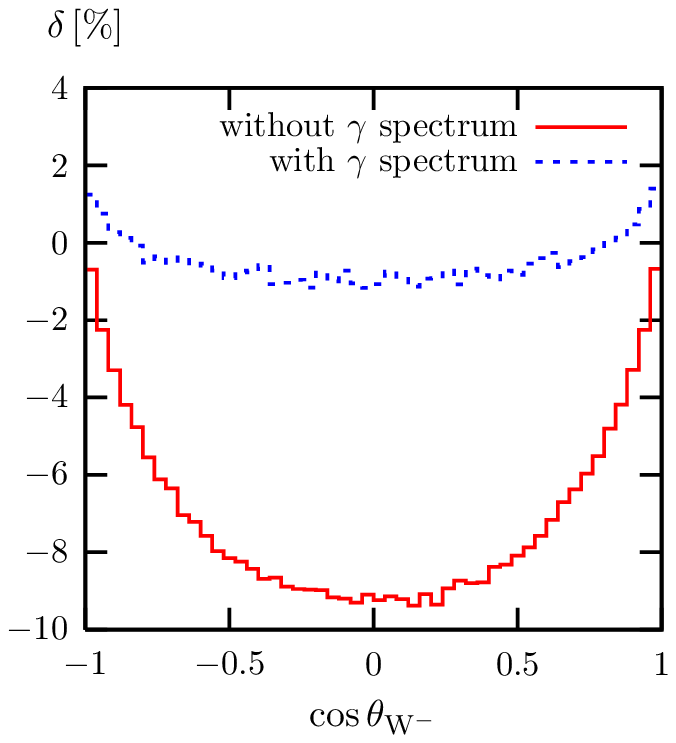}}
\end{picture}}
\vspace*{-1.2em}
\caption{Absolute predictions (upper plots) and relative corrections
(lower plots) for the invariant-mass (l.h.s.) and production-angle 
(r.h.s.) distributions of the 
$\PW^-$ bosons reconstructed from the $\Pd\Pubar$ pairs
in the process $\ga\ga\to\Pne\Pep\Pd\Pubar$ at $\sqrt{s}=500\GeV$
(Taken from \citere{Bredenstein:2005zk}).}
\label{fig:wdist}
\end{figure}

In \reffi{fig:wdist} we show the invariant-mass and production-angle
distributions of the $\PW^-$ boson as recontructed from the
$\Pd\Pubar$ pair in the process $\ga\ga\to\Pne\Pep\Pd\Pubar$,
both with and without convolution over the photon spectrum.
The upper plots show the absolute predictions and the lower plots the
corrections normalized to the Born predictions.
Since we use $\sqrt{s_{\ga\ga}}=500\GeV$ and $\sqrt{s_{\Pe\Pe}}=500\GeV$,
the corrections are shifted upwards when including the photon
spectrum, because the effective energy of the photons is lower.
For the W-invariant-mass distribution,
the shape of the correction, however,
is hardly changed by the convolution over the photon spectrum.
As the shape of the corrections determines a possible shift of the peak of the
invariant-mass distribution, it is of particular importance
in the determination of the W-boson mass. The measurement of the
W-boson mass can, e.g., be used for understanding and calibrating
the detector of a $\ga\ga$ collider.
The distribution in the W-boson production angle
is sensitive to anomalous couplings.
In order to set bounds on these couplings it is mandatory to know
radiative corrections,
because both anomalous couplings and radiative corrections typically
distort angular distributions.
While the correction without the photon spectrum is about $-9\%$
for W bosons emitted perpendicular to the beam, the corrections are
rather small when including the photon spectrum. 
In fact, the relative correction $\de$
is accidentally small at $\sqrt{s_{\Pe\Pe}} \sim 500\GeV$
[cf.\ \reffi{fig:sqrts.spec}(d)]
and might also become larger if other cuts or event selection procedures
are applied.

%




%


\end{document}